\documentclass[showpacs,twocolumn,amsmath,amssymb]{revtex4}
\DeclareMathOperator{\erf}{erf}
\usepackage{graphicx}% Include figure files
\usepackage{dcolumn}% Align table columns on decimal point
\usepackage{bm}% bold math

\begin{document}

\title{Deceleration of continuous molecular beams}% Force line breaks with \\

\author{Eric R. Hudson}
\email{eric.hudson@ucla.edu}
\affiliation{Department of Physics and Astronomy, University of California, Los Angeles, 475 Portola Ave, Los Angeles, CA 90095, USA}%

\begin{abstract}
A method for decelerating a continuous beam of neutral polar molecules is theoretically demonstrated. This method utilizes non-uniform, static electric fields and regions of adiabatic population transfer to generate a mechanical force that opposes the molecular beam's velocity. By coupling this technique with dissipative trap-loading, molecular densities $\geq10^{11}$ cm$^{-3}$ are possible. When used in combination with forced evaporative cooling the proposed method may represent a viable route to quantum degeneracy for a wide-class of molecular species.
\end{abstract}
\maketitle

The electric dipole-dipole interaction between polar molecules is fundamentally different from most interactions between ultracold atoms. While atomic interactions are typically isotropic and comparatively short-ranged, the dipolar interaction is strong, long-range, tunable and anisotropic. This interaction can lead to many novel and exciting phenomena, such as quantum chemistry \cite{roman_v_krems_molecules_2005,hudson_production_2006}, field-linked states \cite{avdeenkov_field-linked_2004}, the possibility for quantum computation \cite{demille_quantum_2002, rabl_hybrid_2006}, and long-range topological order \cite{micheli_toolbox_2006}. Furthermore, the closely spaced, opposite-parity internal levels of molecules, \textit{e.g.} $\Omega$-doublet, rotational, and vibrational levels, present new possibilities for precision measurement of fundamental physics \cite{hudson_cold_2006,flambaum_enhanced_2007,demille_using_2008, demille_[07090963v1]_,zelevinsky_[07081806]_}.

For these reasons, there has been much effort towards developing techniques to produce cold polar molecules. Current molecular cooling techniques can be characterized as either association of ultracold atoms or direct cooling of molecules. While the association of ultracold atoms, either via a Feshbach \cite{ospelkaus_ultracold_2006} or optical resonance \cite{kerman_first_2004}, is capable of producing molecules near quantum degeneracy \cite{Ni_KRb_2008}, these methods restrict experiments to a limited class of molecules -- namely, those composed of laser-cooled atoms. Conversely, direct cooling techniques, such as buffer gas cooling \cite{doyle_buffer-gas_1995} and Stark deceleration \cite{bethlem_decelerating_1999} are capable of producing cold samples from a wide range of polar molecular species. However, despite much work, the molecular density currently attainable via these methods is limited to $\lesssim~10^{7-8}$ cm$^{-3}$. In the case of buffer gas cooling, the molecular density is limited by the presence of the helium buffer gas, which prevents further cooling; for Stark deceleration the poor efficiency ($\sim$0.01\%) of decelerating molecules from a supersonic beam is the limiting factor \cite{Hudson_Efficient_2004}. Furthermore, the `single-shot' nature of these techniques makes it difficult to accumulate cold molecules over time.
\begin{figure}
\resizebox{0.95\columnwidth}{!}{
    \includegraphics{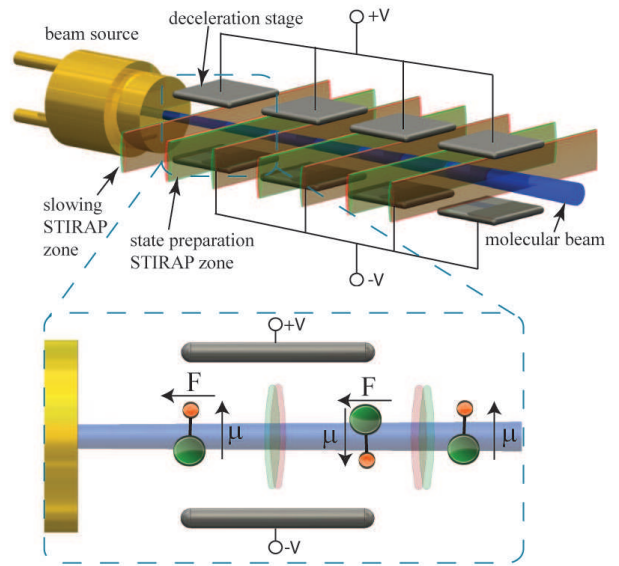}
} \caption{(color online) Schematic of continuous beam decelerator. Weak-field seeking polar molecules exiting the beam source are decelerated as they move into the high electric field region between the electrodes of the deceleration stage. At the stage center, STIRAP is used to transfer the molecules to either a less polarized weak-field seeking state or a strong-field seeking state (inset), resulting in net deceleration of the molecular beam. After the deceleration stage, the molecules are returned to their initial internal state via the inverse STIRAP and the process repeated. For clarity, we have omitted the low-voltage electrodes between the deceleration stages, which are used to maintain molecular orientation. \label{Schematic}}
\end{figure}

Here we propose a new technique for producing continuous beams of cold polar molecules. When coupled with irreversible trap-loading \cite{van_de_meerakker_accumulating_2001,Stuhl_TiOMOT_2008} this continuous beam deceleration technique circumvents the poor efficiency associated with traditional beam deceleration by allowing the accumulation of molecules over many seconds, similar to the loading of a magneto-optical trap (MOT) \cite{Phillips_MOTs_1998}. Moreover, this method appears to allow the accumulation of densities appropriate for evaporative cooling and may represent a viable route to quantum degeneracy for a wide-range of molecular species. In the remainder of this work the method for decelerating continuous molecular beams is detailed and supported with calculations of the overall efficiency.

Because continuous beams are extended in space they are not amenable to traditional Stark decelerators, which rely on well-timed electric field pulses and a spatially compact pulsed molecular beam \cite{bethlem_decelerating_1999}. Figure \ref{Schematic} shows a method of deceleration that instead uses static electric fields, coupled with appropriately placed transition regions, to control the molecular internal state. A molecule in a weak-field seeking state (dipole moment anti-aligned with the electric field) exiting the beam source will be decelerated as it moves into the high electric field region between the electrodes. If at the center of this deceleration stage the molecular internal state is changed to a less polarized weak-field seeking state, the molecule will exit the deceleration stage with less kinetic energy than when it entered. After exiting the stage, the molecule internal state can be returned to its original state and the deceleration process repeated. These transitions can be driven with near unit efficiency by the technique of stimulated Raman adiabatic passage (STIRAP) \cite{bergmann_coherent_1998}.

This approach has several advantages over traditional Stark deceleration. First, this technique allows for continuous guidance of the transverse molecular motion during the deceleration process; transverse motion has recently been shown to be the limiting factor in traditional decelerator efficiency \cite{Sawyer_Mitigation_2008}. Second, as shown in the inset of Fig. \ref{Schematic}, if the  molecule's internal state is instead changed to a strong-field seeking state (dipole moment aligned with the electric field) at the electrodes center, the molecule will be decelerated as it moves out of the high electric field region. Since strong-field seeking states typically experience the largest Stark shift, this technique makes it possible to decelerate a beam of molecules to rest with substantially fewer deceleration stages than required in a traditional decelerator. Third, because molecules can be decelerated in strong-field seeking states and transversely refocussed in weak-field seeking states, this technique can be more efficient for the deceleration of certain heavy molecules than the alternate-gradient deceleration technique \cite{bethlem_alternating_2006}. Fourth, because this technique can decelerate continuous molecular beams it is amenable to the use of a buffer gas beam source \cite{maxwell_high-flux_2005}, which provides a higher molecular flux at a much lower initial beam speed than room temperature molecular beams; this is extremely advantageous since the required number of deceleration stages scales with the square of the molecular beam velocity. Finally, because this technique produces a continuous beam of decelerated molecules it is well-suited for dissipative trap loading, either into a static trap \cite{van_de_meerakker_accumulating_2001} or the recently proposed molecular MOT \cite{Stuhl_TiOMOT_2008}.

\begin{figure}
\resizebox{0.95\columnwidth}{!}{
    \includegraphics{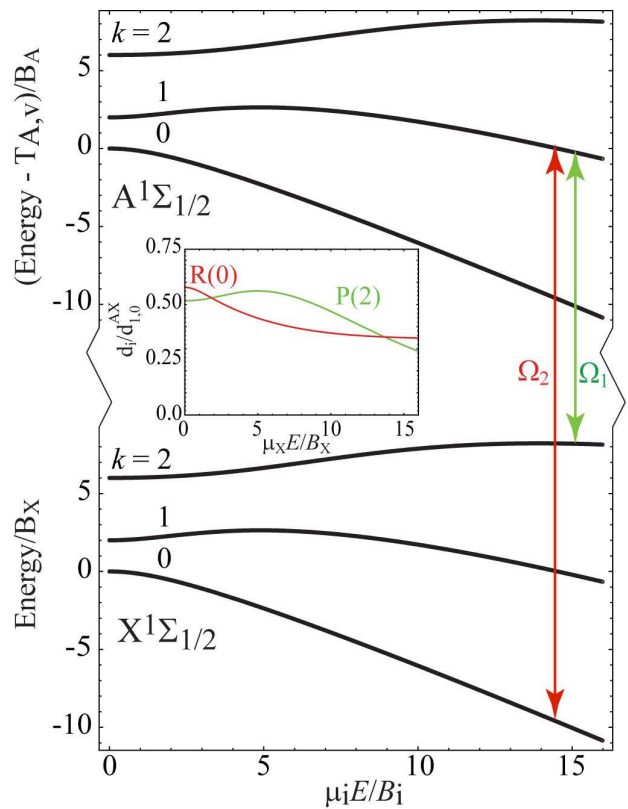}%
} \caption{Stark shift of $^1\Sigma^+$ states, labeled by $k$, and STIRAP transfer scheme. Molecules enter the deceleration stage in the $k=2$ state and exit in $k=0$. If the STIRAP transfer takes place at $\mu_XE/B_i = 15$ then $\sim$3.1 cm$^{-1}$ is removed per deceleration stage. The figure inset shows the relevant H\"{o}nl-London factors for SrO as a function of electric field. \label{StarkShift}}
\end{figure}

For definiteness, consider a molecule with $^1\Sigma^+$ ground and excited states, \text{e.g.} any compound containing one alkaline earth atom and one chalcogen atom. The Hamiltonian for the lowest vibrational level of each electronic state in the presence of an electric field is:
\begin{equation}
H_o = T_{i,v} + B_{i,v}\vec{J}^2 + \vec{\mu}\cdot\vec{E}_{DC},
\label{Hamiltonian}
\end{equation}
$T_{i,v}$ is the vibronic term energy operator with eigenfunctions $\left|\psi_{i,v}\right>$ where $i =$ X,A,..., $v =$ 0,1,2..., and $T_{\rm{X},0} \equiv 0$; $B_{i,v}$ is the rigid-rotor rotational constant for the $i$\textit{th} electronic level with vibrational excitation $v$; $J$ is the total angular momentum of the molecule; $\vec{\mu}$ is the electric dipole moment operator; and $\vec{E}_{DC}$ is the externally applied electric field. The eigenvectors of this Hamiltonian are $\left|\Psi_{i,v,k}\right> = \sum_J a^{i,k}_J\left|J,M\right>\left|\psi_{i,v}\right>$ with energies as shown in Fig. \ref{StarkShift} as a function of electric field. Here the eigenvector label $k$ is equal to the value of $J$ that describes the zero-field wavefunction.

For the deceleration scheme considered here, the $\left|\Psi_{\rm{A},1,1}\right>\leftarrow\left|\Psi_{\rm{X},0,2}\right>$ and $\left|\Psi_{\rm{A},1,1}\right>\rightarrow\left|\Psi_{\rm{X},0,0}\right>$ transitions are used to adiabatically transfer population during the deceleration process. If deceleration by exclusively weak-field seeking molecules is desired (for transverse guiding purposes) the $\left|\Psi_{\rm{X},0,0}\right>$ state can be replaced by the $\left|\Psi_{\rm{X},0,1}\right>$ state. Note that a two-photon transition between states with $\left|\Delta J\right|$ = 1 is possible here because the static electric field mixes states of opposite parity. In the rotating frame, the Schr\"{o}dinger equation for the evolution of the population of the relevant molecular states is given as \cite{bergmann_coherent_1998}:

\begin{equation}\label{SchroEqnTransform}
i\frac{d}{dt}\left[
  \begin{array}{c}
    c^X_{0,2}\\
    c^A_{1,1}\\
    c^X_{0,0}\\
  \end{array}
\right] =
 \left[
  \begin{array}{ccc}
    0 & -\frac{\Omega_1(\rm{t})}{2} & 0 \\
    -\frac{\Omega_1(\rm{t})}{2} & -i\frac{\Gamma}{2} & -\frac{\Omega_2(\rm{t})}{2}\\
    0 & -\frac{\Omega_2(\rm{t})}{2} & \Delta \\
  \end{array}
\right] \left[
  \begin{array}{c}
    c^X_{0,2}\\
    c^A_{1,1}\\
    c^X_{0,0}\\
  \end{array}
\right],
\end{equation}
where $\Gamma$ is the natural linewidth of the A state in radians per second, $\Omega_n(t) = \frac{\rm{d}_n\emph{E}_n(t)}{\hbar}$ with $\emph{E}_n(t)$ the envelope of the laser's electric field amplitude, and $\left|c^i_{v,k}\right|^2$ is the probability of the molecule being found in the state $\left|\Psi_{i,v,k}\right>$. The transition dipole moments are given in terms of the dipole operator, d, as
\begin{equation}
\rm{d}_n = \left<\Psi_{\rm{A},1,1}\right|\rm{d}\left|\Psi_{\rm{X},0,2(2-n)}\right>,
\label{DipoleMoment}
\end{equation}
which can be calculated from the relevant transition moments: $\left<\psi_{\rm{A},v}\right|\left<J+1,M\right|\rm{d}\left|JM\right>\left|\psi_{\rm{X},v}\right> = \rm{d}^{\rm{AX}}_{v',v} \sqrt{(J+M+1)(J-M+1)/((2J+1)(2J+3))}$ and $\left<\psi_{\rm{A},v}\right|\left<J-1,M\right|\rm{d}\left|JM\right>\left|\psi_{\rm{X},v}\right> = \rm{d}^{\rm{AX}}_{v',v} \sqrt{J^2-M^2/((2J+1)(2J-1))}$, where $\rm{d}^{\rm{AX}}_{v',v}$ contains both the electronic and vibrational contributions to the transition dipole moment.

As a specific example, Eqns. \ref{Hamiltonian} and \ref{SchroEqnTransform} have been solved for the strontium monoxide molecule, SrO, ($B_X = 0.33798$~cm$^{-1}$, $\mu_X = 8.9$~D, $B_A = 0.3047$~cm$^{-1}$ \cite{radzig_reference_1985}, $\mu_A \approx 2.5$~D \cite{AstateDipoleNote}, $\Gamma$ = 2$\pi\times$3.7~MHz \cite{jessie-thesis}, and $\rm{d}^{\rm{AX}}_{1,0} = 0.43$~D \cite{Nicholls_FC_factors}) in a 30~kV/cm electric field ($\mu_X E/B_X \approx 15$). For a geometry similar to Ref. \cite{nguyen_efficient_2000}, with laser beams of axial and transverse $e^{-2}$ intensity waist $w_A\times w_\perp$ of 1$\times$10 mm$^2$ and total power of 1 W, the transfer efficiency has been calculated to be $\geq$~90\% throughout the deceleration process -- because the pulse area scales as $v^{-1}$, the transfer efficiency increases with deceleration stage number, $N$, as $N^{1/2}$. At this field, and with transfer from the $\left|2,0\right>$ to $\left|0,0\right>$ state at the stage center, $\sim$3.1 cm$^{-1}$ is removed from the molecular kinetic energy at each deceleration stage, allowing a 100 m/s beam of SrO to be decelerated to rest with only twenty deceleration stages.

\begin{figure}
\resizebox{0.95\columnwidth}{!}{
    \includegraphics{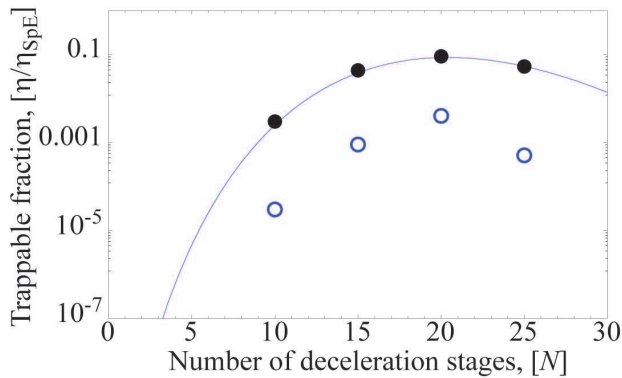}%
} \caption{Results of three-dimensional Monte Carlo simulation of the deceleration process. The fraction of molecules that exit the decelerator and can be coupled into the trap is displayed versus the number of deceleration stages used. Open circles are for a traditional buffer gas beam of SrO molecules ($T_\perp = 4$ K). Closed circles are for a molecular beam that has undergone transverse laser cooling ($T_\perp = 10~\mu$K). The line represents the maximum trappable fraction as predicted by Eq. \ref{TrapFractionEqn}.\label{TrappedFraction}}
\end{figure}

If the decelerated molecules are loaded into the microwave trap of Ref. \cite{DeMille_Microwave} via a spontaneous emission loading scheme similar to that of Ref. \cite{van_de_meerakker_accumulating_2001}, then molecules exiting the decelerator with a final velocity less than the maximum trappable velocity, $v_{\rm{trap}}$, can be accumulated in the trap. Since, all molecules lose the same amount of energy $\Delta E$ per deceleration stage these trappable molecules enter the decelerator with a velocity in the range \{$\sqrt{\frac{2N\Delta E}{m}},\sqrt{v_{\rm{trap}}^2 + \frac{2N\Delta E}{m}}$\}. Assuming the buffer gas beam source can be described by the longitudinal velocity distribution,
\begin{equation}
f(v) = \frac{1}{\Delta v}\sqrt{\frac{\ln(2)}{\pi}}
\exp{\left[-\left(\frac{v-v_0}{\Delta v/\sqrt{\ln 2}}\right)^2\right]},
\label{LongitudinalVelocityDistribution}
\end{equation}
the fraction of trapped molecules is then given as
\begin{eqnarray}
\eta = &\frac{\eta_{\rm{S.E.}}\eta_{\perp}}{2}\left(\erf\left[\frac{\sqrt{v_{\rm{trap}}^2 + \frac{2N\Delta E}{m}}-v_0}{\Delta v/\sqrt{\ln 2}}\right]-\erf\left[\frac{\sqrt{{2N\Delta E}{m}}-v_0}{\Delta v/\sqrt{\ln 2}}\right]\right),\nonumber\\
&\label{TrapFractionEqn}
\end{eqnarray}
where $\eta_{\rm{S.E.}}$ is the efficiency of the spontaneous emission trap loading \cite{van_de_meerakker_accumulating_2001} and $\eta_{\perp}$ is the fraction of molecules that remain within the transverse bounds of the decelerator during the deceleration process \cite{Hudson_Efficient_2004}. For loading of the microwave trap by excitation of the $\left|\Psi_{A,1,0}\right>\leftarrow\left|\Psi_{X,0,2}\right>$ transition followed by spontaneous emission on the $\left|\Psi_{A,1,0}\right>\rightarrow \left|\Psi_{X,0,0}\right>$ transition \cite{Amar_TrapLoading}, and assuming excitation saturation, Eq. \ref{DipoleMoment} yields $\eta_{\rm{S.E.}} = 0.4$ \cite{TrapLoadingNote}. For reference, Eq. \ref{TrapFractionEqn} is plotted in Fig. \ref{TrappedFraction} for SrO ($v_0$ = 100 m/s and $\Delta v$ = 25 m/s) and the microwave trap parameters of Ref. \cite{DeMille_Microwave} ($v_{\rm{trap}} = 2$ cm$^{-1}$).

The transverse efficiency $\eta_{\perp}$ depends sensitively on the details of the decelerator design and molecular beam, and can only be estimated from detailed Monte Carlo trajectory simulations. Since most of the deceleration takes place in a strong field seeking state and the strongest fields are near the electrode surface, molecules are defocused transversely as they are decelerated. Thus, a mechanism for transverse refocusing of the molecular beam must be added to the decelerator to prevent complete loss of the beam due to transverse defocussing. The techniques of transversely focusing molecular beams are the subject of much on-going work \cite{Bas_Transverse, Sawyer_Mitigation_2008} and a complete optimization of the beamline is beyond the scope of the present work. Nonetheless, Monte Carlo trajectory simulations were performed for a decelerator geometry with a hexapole focusing element \cite{Anderson_Hexapole_1997} ($r_o$ = 1 cm and $V_o$ = 35~kV) added after every fifth deceleration stage. The results of these simulations are shown in Fig. \ref{TrappedFraction} as open circles for a beam of SrO molecules with $v_0$ = 100~m/s, $\Delta v$ = 25~m/s, and transverse beam temperature $T_\perp$ = 4~K. At twenty stages of deceleration, $\eta$ reaches a maximum of $\sim$0.002. Thus, with an initial molecular beam flux of $10^{13}$ molecules/s \cite{maxwell_high-flux_2005} and trap volume of $V \approx \frac{4}{3}\pi$ cm$^{-3}$ \cite{DeMille_Microwave}, the trap density accumulates at a rate of $K \approx 5\times10^{9}$ cm$^{-3}$s$^{-1}$. For a background gas limited lifetime of $\tau \approx 100$ s, an average trap density of $\rho = K\tau \approx 5 \times 10^{11}$ cm$^{-3}$ is expected.

A recent proposal has shown how certain molecules may be laser-cooled \cite{Stuhl_TiOMOT_2008} and collected in a MOT. While most molecules are not amenable to the molecular MOT cooling scheme, many molecules possess relatively diagonal Franck-Condon factors allowing the possibility of scattering many photons before optical pumping to undesired states occurs. Thus, cooling only the transverse degree of freedom of a molecular beam, which requires the scattering of a maximum of a few thousand photons, may be possible for many molecular species. For reference, simulations were also performed for a SrO molecular beam that has undergone transverse laser cooling to $T_\perp$ = 10 $\mu$K. As shown in Fig. \ref{TrappedFraction} as closed circles, the reduced transverse beam temperature makes it possible to recover the maximum trapped fraction as predicted by Eq. \ref{TrapFractionEqn}. Of course, it may be possible to achieve the maximum efficiency, even for $T_\perp$ = 4 K, with more elaborate transverse guiding schemes (\textit{e.g.} addition of more focusing elements, different order focusing elements, voltage scaling \cite{Sawyer_Mitigation_2008}) and this is the subject of on-going work.

An isolated dilute gas can be efficiently cooled by self-accelerating evaporative cooling as long as the product of the elastic collision rate and trap lifetime exceeds $\sim$150 \cite{davis_analytic_1995}. For polar molecules in their absolute ground state, the collision rate is expected to be accurately predicted by the Eikonal approximation for T $\gtrsim$ 200~fK \cite{DeMille_Microwave, ScatteringRateNote} and is given as \cite{Bohn_dipole_2009} as $\gamma = \rho\sigma v = \frac{2 \rho \mu^2}{3\hbar \epsilon_o} \approx 2\times10^5~\rm{Hz}.$ This large scattering rate (by atomic physics standards) is due to the long-range nature of the dipolar interaction between polar molecules in their absolute ground state. Since $\gamma\tau \approx  10^7$, the criteria for efficient evaporation is clearly satisfied and evaporative cooling appears to be a viable route to quantum degeneracy. However, the anisotropic dipolar interaction may lead to a significantly peaked differential cross-section, which, due to the reduced momentum transfer for small angle scattering, could reduce the efficiency of the cooling below this estimate. Furthermore, species dependent, density limiting effects such as three-body recombination have not been studied in detail and could lead to unexpected losses; in analogy to evaporative cooling of cesium it is likely that an externally applied electric or magnetic field can be used to mitigate the effect.

Because the production of the necessary electric fields and ultra-high vacuum is now routine, the main technical challenge in constructing the continuous beam decelerator is the required voltage stability. Since the two-photon detuning, $\Delta$, depends on the electric field, any change in the field during the deceleration process diminishes the efficiency of the STIRAP process. At 30 kV/cm the energy dependence of the $\left|20\right>\left|\psi_{\rm{X},v}\right>$ state on electric field is weak compared to the field dependence of the  $\left|00\right>\left|\psi_{\rm{X},v}\right>$ state. Therefore, $\Delta$ depends on electric field as $\Delta \approx \left< \mu_0 \right>\delta |\vec{E}|$, where $\left< \mu_0 \right>$ is the ground rotational state, laboratory-frame dipole moment (6.9 D for SrO \cite{DeMille_Microwave}) and $\delta |\vec{E}|$ is any drift of the electric field during the deceleration process. Maintaining maximum STIRAP efficiency requires $\Delta \lesssim $ $2\pi v/w_A$. Since the STIRAP efficiency is $\propto \sqrt{w_A}/v$, while the pulse time is $\approx w_A/v$, the beam waist may be decreased with the square of the velocity to maintain maximum transfer efficiency. Thus, the tightest constraint on field stability comes from the first deceleration stage, which in the configuration of Fig. \ref{StarkShift} leads to $\Delta \lesssim  2\pi\times$100 kHz, or a required electric field stability of $\delta |\vec{E}|/|\vec{E}| \sim 10^{-6}$. For instance, a 1 cm electrode spacing requires a challenging, but possible, voltage stability of 0.1 V out of 30 kV. Any inhomogeneity between deceleration stages may be offset by local voltage adjustments. Moreover, it may be possible to \textit{utilize} a field inhomogeneity to produce an adiabatic rapid passage transfer \cite{kasuga_adiabatic_1976}, greatly reducing the constraints on field stability.

In summary, a general technique for producing cold polar molecules has been described. This method promises large improvement over other direct cooling methods by allowing deceleration of a continuous molecular beam as opposed to the single-shot nature of current techniques. By coupling the decelerated molecular beam into a trap, densities $\ge10^{11}$ cm$^{-3}$ seem attainable, which represent an improvement of up to 4 orders of magnitude over current technology. At these densities it appears efficient evaporative cooling to quantum degeneracy may be possible. Finally, in this work STIRAP has been used to accomplish the necessary state transfer, however, in the presence of an electric field, parity is mixed and a single photon can transfer population between the relevant levels. Thus, it may be possible to perform direct transfer by THz radiation, further simplifying the design.

\acknowledgements{The author thanks D. DeMille, W.G. Rellergert, and J. Ye for useful comments.}
%\bibliography{STIRAPSlowingBib}% Produces the bibliography via BibTeX.

\begin{thebibliography}{38}
\bibitem[{Tra()}]{TrapLoadingNote}
\bibinfo{note}{Strictly speaking, this value is the maximum of
  $\eta_{\rm{S.E.}}$ since some molecular trajectories may not be coupled into
  the trap region. However, with proper transverse focusing this value can be
  closely approached.}
\bibitem[{Sca()}]{ScatteringRateNote}
\bibinfo{note}{As the molecules are cooled this rate will eventually reach the
  unitarity limit where it will become a constant. However, because dipolar
  scattering is very long-range this does not affect the result.}
\end{thebibliography}

\end{document}